\documentclass[runningheads]{llncs}
\usepackage{amsmath, amsthm, amssymb}
\usepackage{graphicx}
\usepackage{caption}
\usepackage{subcaption}
\usepackage{color}
\usepackage[english]{babel}
\usepackage{multirow}
\usepackage{boldline}
\usepackage{hyperref}
\usepackage[dvipsnames]{xcolor}
\usepackage{float}
\usepackage{amssymb}
\usepackage{pifont}
\newcommand{\emoji}[1]{\includegraphics[width=1em]{#1}}

\newcommand{\brains}{\emoji{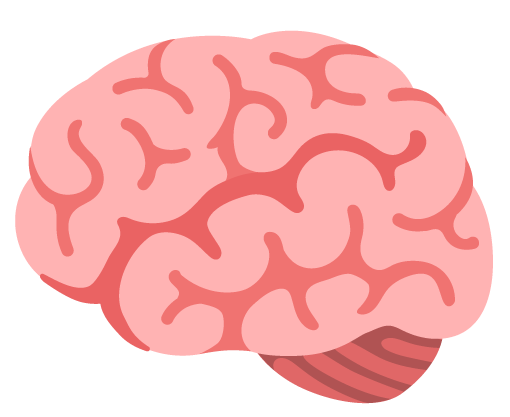}BRAINS-45K}
\newcommand{\brainsshort}{\emoji{brain_cropped.png}-45K}
\newcommand{\framework}{AMAES}
\newcommand{\yucca}{Yucca}
\newcommand{\dplus}[1]{\;\text{\scriptsize{\color{ForestGreen} +#1}}}
\newcommand{\dminus}[1]{\;\text{\scriptsize{\color{red} -#1}}}
\newcommand{\hlinebold}{\specialrule{.1em}{.05em}{.05em}}

\usepackage{ctable}

\begin{document}

\title{\framework: Augmented Masked Autoencoder Pretraining on Public Brain MRI Data for 3D-Native Segmentation}
\titlerunning{Augmented Masked Autoencoder Pretraining for 3D Brain Segmentation}
\author{Asbjørn Munk\inst{1,2*} \and
Jakob Ambsdorf\inst{1,2*} \and
Sebastian Llambias\inst{1,2} \and
Mads~Nielsen\inst{1,2}
}
% \author{***}
%
\authorrunning{A. Munk et al.}
%\authorrunning{***}
\institute{Department of Computer Science, University of Copenhagen, Copenhagen, Denmark \and
Pioneer Centre for AI, Copenhagen, Denmark \\
\email{\{asmu,jaam\}@di.ku.dk}\\
*Equal contribution
}
% \institute{*** \\ \email{***}}
%
\maketitle              % typeset the header of the contribution
\begin{abstract}
This study investigates the impact of self-supervised pretraining of 3D semantic segmentation models on a large-scale, domain-specific dataset. We introduce \brains, a dataset of 44,756 brain MRI volumes from public sources, the largest public dataset available, and revisit a number of design choices for pretraining modern segmentation architectures by simplifying and optimizing state-of-the-art methods, and combining them with a novel augmentation strategy. The resulting AMAES framework is based on masked-image-modeling and intensity-based augmentation reversal and balances memory usage, runtime, and finetuning performance. Using the popular U-Net and the recent MedNeXt architecture as backbones, we evaluate the effect of pretraining on three challenging downstream tasks, covering single-sequence, low-resource settings, and out-of-domain generalization. The results highlight that pretraining on the proposed dataset with AMAES significantly improves segmentation performance in the majority of evaluated cases, and that it is beneficial to pretrain the model with augmentations, despite pretraing on a large-scale dataset. Code and model checkpoints for reproducing results, as well as the \brains~dataset are available at \url{https://github.com/asbjrnmunk/amaes}.

\end{abstract}
\section{Introduction}
\begin{figure}[t]
    \centering
    \includegraphics[width=\textwidth]{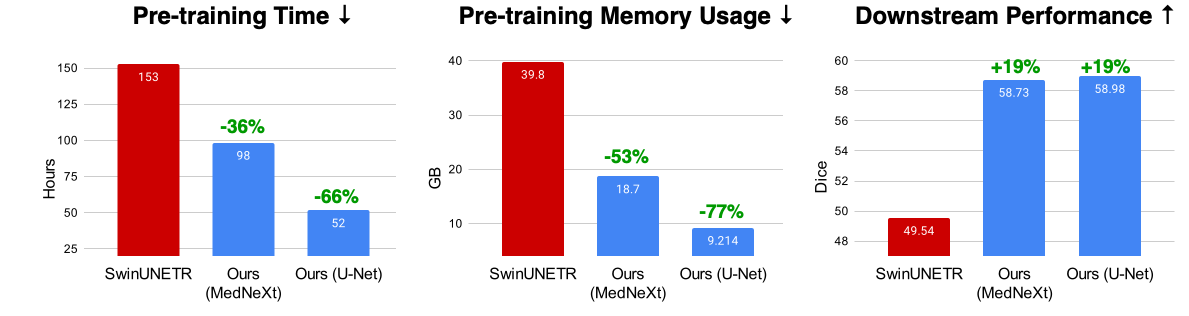}
    \caption{\framework~provides \textit{efficient} 3D pretraining for segmentation networks requiring less resources than SwinUNETR while improving on downstream performance. Downstream performance is on the BraTS21 dataset, see Section \ref{sec:results}. The MedNeXt model is MedNeXt L (55 mio. parameters), the U-Net is U-Net XL (90 mio. parameters). SwinUNETR has 60 mio. parameters. Memory usage is recorded with a batch size of two for all models. All results were obtained using Nvidia H100 GPUs and with mixed 16-bit precision using uncompiled models.}
    \label{fig:three graphs}
\end{figure}

% In recent years, large scale self-supervised pretraining has completely revolutionized NLP and Computer Vision.
Large-scale self-supervised pretraining promises immense potential for medical image segmentation. Works in computer vision from the natural image domain demonstrate that state-of-the-art results are achieved by scaling model capacity and pretraining on large datasets, enabling increased generalization performance, few- or zero-shot predictions, and highly semantic embeddings~\cite{he2022masked}. 
%\ja{talk about contrastive and MIM here.}
% Currently two streams 
To enable self-supervised representation learning in vision, various methods have been proposed. Early works involved reversing image degradations, such as denoising images~\cite{vincent2008extracting}, solving jigzaw puzzles \cite{noroozi2016unsupervised}, or re-colorizing images~\cite{zhang2016colorful}.
Masked-Image-Modeling (MIM) proved to be an especially successful and scalable variant, where patches of the input image are masked and learned to be reconstructed either directly or indirectly by predicting an embedding~\cite{bao2021beit,he2022masked}. A Masked Autoencoder (MAE) is an encoder-decoder architecture that reconstructs the entire image from a masked version~\cite{he2022masked}.
Another family of approaches applied contrastive learning, where semantically similar transformed inputs are mapped closely in a latent space, while dissimilar inputs are pushed apart~\cite{chen2020simple}. Self-distillation is based on a related idea, enforcing similar embeddings for semantically close inputs for a student and teacher network~\cite{grill2020bootstrap}.
Recent approaches combined these image-level contrastive or self-distillation objectives with patch-level MIM to enhance the spatial representation performance for downstream segmentation tasks~\cite{bao2021beit,tang2022swinunetrpret}.

Evidence from medical vision indicates that the benefits of pretraining are transferable to biomedical image segmentation~\cite{chaitanya2020contrastive}. Yet, the current method of choice for solving segmentation tasks is still a \textit{fully-supervised} paradigm~\cite{isensee2024nnu}. Due to a heavily optimized data augmentation pipeline, the fully-supervised approach can be effective even with limited labeled data \cite{chlap2021review,Isensee2020nnUNet}. However, to fully leverage the potential of large-scale models in medical image segmentation as witnessed in computer vision, more training data is required, and the supervised approach is inherently limited by labeling cost. Large-scale \textit{unlabeled} medical imaging datasets could resolve this need, but they are not readily available to the public, hindering progress for methodological research on how to harness unlabeled data for downstream segmentation most effectively. To address the need for pretraining datasets in the brain MRI domain, we propose \brains, to the best of our knowledge the largest public pretraining dataset for brain MRI to date\footnote{\brains~ is an order of magnitude larger than the pretraining dataset used in SwinUNETR, the current state-of-the art methodology.} comprising of 44,756 volumes collected from public sources.

Using this large-scale dataset, we are proposing the AMAES (Augmented Masked Auto Encoder for 3D Segmentation) framework for pretraining modern semantic segmentation architectures. AMAES combines the successful Masked Image Modeling (MIM) task~\cite{he2022masked} with augmentation reversal, both of which can be viewed as special cases of denoising autencoding ~\cite{vincent2008extracting}, with the goal of learning local, detailed representations that are invariant to intensity shifts between different domains. The resulting framework balances factors such as GPU memory usage, pretraining runtime, and downstream task performance. We compare AMAES against the state-of-the-art pretraining method proposed for SwinUNETR~\cite{tang2022swinunetrpret}, which combines MIM with a rotation prediction task and a contrastive learning objective. We find that in practice, solving the rotation problem is trivial while the contrastive loss adds little benefit provided the relatively small batch sizes possible in 3D imaging. Our simplified framework results in a significant speedup and memory reduction while improving downstream performance.

We evaluate our method using two backbone architectures: (i) The U-Net, as the most popular semantic segmentation model, and (ii) MedNeXt, as a recently proposed modernization. The \brains~dataset and proposed AMAES framework for pretraining segmentation models contribute to building a foundation for the next generation of large-scale biomedical image segmentation models.

\begin{table}[t]
\caption{\textbf{The \brains~dataset.} The dataset is of high diversity and contains a wide range of different sequences, combining data acquired at both 1.5T and 3T at multiple spatial resolutions. All data is preprocessed to 1mm isotropic spacing, and intensities are normalized to the $[0,1]$ interval.}
\resizebox{\textwidth}{!}{
\begin{tabular}{l|llll}
\hlinebold
\textbf{Source}               & \textbf{Subjects}\; & \textbf{Sessions}\; & \textbf{Volumes}\; & \textbf{Sequences} \\ 
\hline
ADNI \cite{adni}              & 1563                & 8019                &  9548              & T1, Flair, T2* \\
OASIS3 \cite{oasis3}          & 1376                & 2839                & 15872              & T1, T2, Flair, TOF, DWI, T2*, SWI \\
OASIS4  \cite{oasis4}         & 661                 & 674                 & 4021               & T1, T2, Flair, T2*, SWI \\
PPMI  \cite{ppmi}             & 2011                & 2979                & 9554               & T1, T2, Flair, DWI, T2* \\
BraTS21 (train) \cite{brats21} & 938                 & 938                 &  3752              & T1, T2, Flair, T1ce \\
ISLES2022 \cite{isles22} & 250                 & 250                 & 750                & DWI, Flair, ADC \\
WMH (train) \cite{wmh}        & 60                  & 60                  & 120                & T1, Flair \\ 
MSSEG1 (train) \cite{msseg1}  & 15                  & 15                  & 75                 & T1, T2, Flair, T1ce \\ 
MSD BrainTumor (train) \cite{msd} & 266                 & 266                 &  1064              & T1, T2, Flair, T1ce \\
\hline
\textbf{Total}                & \textbf{7140}      & \textbf{16040}       & \textbf{44756}  \\
\hlinebold
\end{tabular}
}
\label{tab:dataset}
\end{table}

\section{\brains~dataset}

We propose \brains~ a dataset of $44.756$ Brain MRI volumes collected from public sources, featuring a diverse set of acquisition parameters and patient populations. The dataset is a compilation of data from four large non-labelled datasets (ADNI \cite{adni}, OASIS3 \cite{oasis3}, OASIS4 \cite{oasis4}, PPMI \cite{ppmi}) and five challenge datasets (MSD \cite{msd}, BraTS21 \cite{brats21}, ISLES22 \cite{isles22}, WMH \cite{wmh}, MSSEG1 \cite{msseg1}). The resulting dataset is highly heterogeneous and assembled to simulate the characteristics of clinical data. Overview of the dataset can be found in Table~\ref{tab:dataset}. The dataset includes data acquired at both 1.5T and 3T. PPMI data is openly available from PPMI. ADNI data used in preparation of this article were obtained from the Alzheimer’s Disease Neuroimaging Initiative (ADNI) database. To compile the diverse data into a suitable pretraining dataset, we transformed all volumes to the RAS coordinates system, resampled to isotropic 1mm spacing, clipped values to the 99\% percentile, and $z$-normalized on a volume level. Afterwards, each volume is cropped to the minimum bounding box of the brain. The preprocessing pipeline is based on the \yucca~framework for medical imaging \cite{llambias2024yucca}. Code to reproduce the final version of the dataset is provided in the project repository. \textbf{Usage of the dataset needs to comply with the individual legal requirements of each source.}

\section{Method}
\begin{figure}[t]
    \centering
    \includegraphics[width=\textwidth]{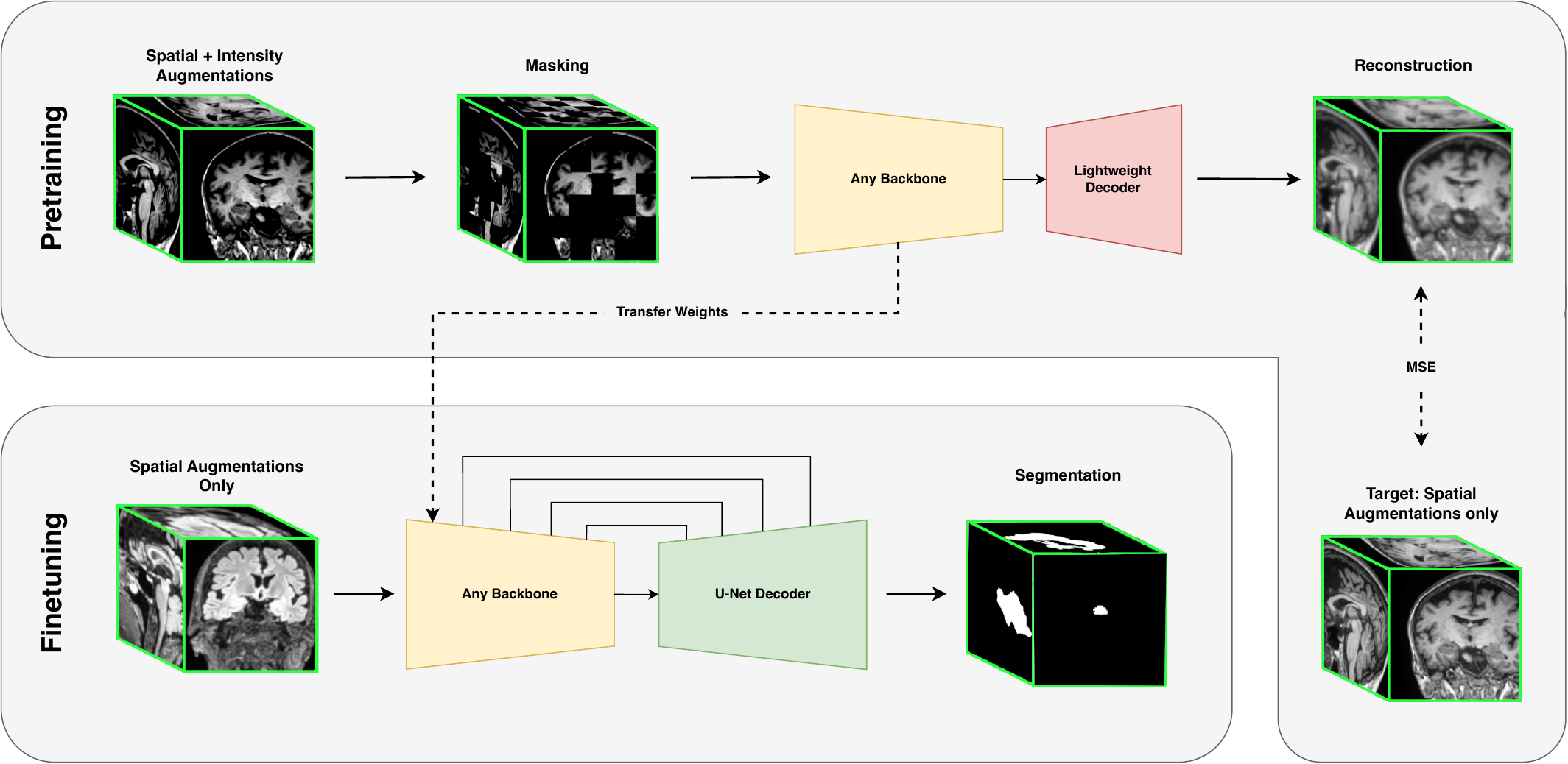}
    \caption{\textbf{Graphical overview of the \framework~framework}. During pretraining, spatial and intensity-based augmentations are applied to an image patch. The patch is masked and passed through the model, which consists of a backbone encoder and a lightweight decoder, to reconstruct the image. The reconstruction target is the unmasked image, with \textit{only} spatial transformations applied. During finetuning, only spatial augmentations are applied to the input. The backbone encoder weights are transferred, while a new U-Net decoder is initialized. Skip-connections are only used during finetuning.}

    \label{fig:architecture}
\end{figure}

In this section we briefly describe \framework. Our approach is highlighted in Figure~\ref{fig:architecture}. The overall goal of the framework is to provide an effective and efficient pretraining strategy for large-scale datasets, requiring a balance between memory usage, runtime and downstream performance (see Figure~\ref{fig:three graphs}).

\subsection{Pretraining}
\label{subsec:pretrain}
\framework~ is based on a Masked Autoencoder (MAE) architecture with a light weight decoder and a specific augmentation scheme. Importantly, the architecture assumes no specific choice of encoder and can be used with any encoder that is compatible with the skip connections of the U-Net decoder.

\textbf{Pre-text task.} Prior work has centered around 2D natural images \cite{Woo2023ConvNeXtV2,he2022masked}, however, previous work showed that a Masked Autoencoder works well in 3D by applying it to 3D CT Brains using the SwinUNETR architecture~\cite{tang2022swinunetrpret}. The SwinUNETR pretraining objective further included both a rotation and a contrastive loss. However we find that the rotation task was near-trivial for the models to solve and thus provided no tangible utility, while the contrastive loss resulted in the memory footprint during pretraining to be doubled, without significant performance gains (see Appendix~\ref{app:contrastive-rotation}). Threfore, we reduce the pre-text task for \framework~to masked reconstruction, and instead increase the batch size and double the number of training epochs. For each 3D input patch, the input is divided into subpatches of size $4\times4\times4$ voxels, and each subpatch is masked with probability 0.6, following prior work \cite{Woo2023ConvNeXtV2}.

\begin{table}[t]
\centering
\caption{\textbf{Choice of decoder.} By choosing a light weight decoder we both \textit{increase} downstream performance and speed up training by 25\%. %The light weight decoder is a U-Net-style decoder using \textit{single} blocks of upsampling, convolution, dropout, and instance normalization.
Finetuning Dice with one standard deviation is reported as 6-fold cross validation on BraTS21 with 20 training samples \textit{on the validation split ($n_\text{cv} = 866$)}.
}
\begin{tabular}{llccc}
\hlinebold
Encoder backbone       & Decoder                & \begin{tabular}[c]{@{}c@{}}Finetuned\\ \textit{dice}\end{tabular} & \begin{tabular}[c]{@{}c@{}}Pretraining runtime\\ \textit{hours}\end{tabular}  & Speedup \\ \hline
\multirow{2}{*}{U-Net Encoder} & Standard U-Net decoder & $56.65 \;\pm1.85$ & $39$ & - \\
                       & Light-weight decoder   & $58.61 \; \pm1.94$ & $31$ & $\mathbf{1.25\times}$  \\ \hlinebold
\end{tabular}
\label{tab:decoder_choice}
\end{table}

\textbf{Decoder architecture.} Since the decoder is discarded after pretraining, it is desirable to use as small a decoder as possible. This not only decreases training time and memory usage, but also forces the network to learn expressive representations in the encoder.
% as expressive a representation as possible. 
We find that a light weight version of the standard U-Net decoder provides a good balance between flexibility, memory efficiency and parameter count. Table \ref{tab:decoder_choice} shows results of pretraining with two versions of the U-Net decoder: the standard 3D U-Net decoder and one with half the number of convolutional layers (denoted light weight decoder). Interestingly, the light weight decoder not only decreases pretraining time with 20\%, but also further increases finetuning performance by $1.96$ dice points.

\textbf{Skip connections.} While skip connections are used during finetuning, our proposed encoder-decoder architecture does not include skip connections during pretraining. We found that using skip connections during pretraining tended to slightly adversely affect downstream performance, with a finetuning dice score of $56.22$ with and $56.65$ without, on 6-fold cross-validation on the BraTS21 dataset. While the difference is not significant, we choose to not use skip connections in the pursuit of the simplest possible setup.

\textbf{Augmentations.} To learn representations which are robust to changes in the input domain, we apply an extensive augmentation pipeline to the data during pretraining. The full list of augmentations include all augmentations used in nnUnet \cite{Isensee2020nnUNet} and is given in Appendix \ref{app:augmentations}. We want our representations to be \textit{covariant} to any spatial augmentations (rotation, scaling), while we seek to learn representations that are \textit{invariant} to all other augmentations, for instance additive noise. We thus only apply spatial augmentations to the reconstruction target. This is visualized in Figure~\ref{fig:architecture}. Table~\ref{tab:augmentation} shows the result of applying augmentations to the pretraining data: when applying augmentations during pretraining, we see a $1.24$ Dice point increase in finetuning performance, compared to no augmentations, when finetuned with only spatial augmentations.

\subsection{Finetuning}
\label{subsec:finetune}
For finetuning, we transfer the backbone encoder weights from pretraining and attach a new standard U-Net decoder with skip connections between the encoder and the decoder. We only use spatial augmentations (rotation, scaling) during finetuning, since this setup performed better empirically. Upon closer inspection, the learning curves showed signs of catastrophic forgetting during the second half of training when finetuned with the full augmentation package.

\begin{table}[t]
\centering
\caption{\textbf{Choice of augmentation scheme.} Best results are achieved by pretraining such that the model is \textit{invariant} to intensity-based augmentations and \textit{covariant} to spatial augmentations (rotation, scaling). Finetuning Dice with one standard deviation is reported as 6-fold cross validation on BraTS21 with 20 training samples \textit{on the validation split ($n_\text{cv} = 866$).}}.
\begin{tabular}{lccc}
\hlinebold
 Backbone\: & Pretraining\: & \multicolumn{1}{c}{\begin{tabular}[c]{@{}c@{}}Finetuning\\ \textit{dice}\end{tabular}} \\ \hline
\multirow{2}{*}{U-Net} & No Augmentations  & $58.61 \; \pm 1.94$                                                                   \\
                       & All Augmentations & $59.85 \; \pm 0.98$                                                                   \\ \hlinebold
\end{tabular}
\label{tab:augmentation}
\end{table}

\section{Experimental Design}
\subsection{Datasets}
We evaluate \framework~ on three different segmentation tasks covering tumor (BraTS21 \cite{brats21}), ischemic stroke lesion (ISLES22 \cite{isles22}) and white matter hyperintensity (WMH \cite{wmh}). To test the models few-shot learning ability, we evaluate the models in a low resource setting: we only use $n_\text{train} = 20$ labeled examples for finetuning, while we evaluate the models on the remaining data. We choose $n_\text{train} = 20$ to realistically model what one medical expert can manually label with limited effort. Since the pretraining dataset considers each sequence individually, we evaluate the model's ability to perform downstream tasks in a single sequence setup, and thus for BraTS21 and WMH we consider only the flair modality, for ISLES22 we consider only the DWI modality.

Hyperparameters were selected on $3$-fold cross-validation splits of the BraTS21 dataset. We do not tune hyperparameters individually for each fold or for each dataset. All results are averages over $6$ folds. For BraTS21, we evaluate the model on a 20\% holdout set. For ISLES22 we follow \cite{Isensee2024nnUnetRevisited} and provide results from the rotating cross-validation splits. For WMH we both perform cross-validation evaluation (denoted \textit{"WMH"}) and test set evaluation. The test set contains data from scanners not present in the training set, which are also not present in \brains, and thus evaluates the models ability to generalize to unseen data. We call this dataset "\textit{WMH OOD}" (out-of-distribution). 

\subsection{Configurations}
Our baselines, pretraining and finetuning are implemented using PyTorch, PyTorch Lightning, and the \yucca~ library for medical image analysis \cite{llambias2024yucca} as a back-bone. All models are trained using 16-bit mixed-precision to limit memory footprint. Models were trained patch-based, using an input patch size of $128 \times 128 \times 128$. Models are pretrained for $100$ epochs (one epoch is one patch from each volume) and trained supervised for $2500$ steps on downstream datasets. The exact hyperparameters used for each model is given in Appendix \ref{app:hyperparameters}. All models are pretrained on two H100 GPUs, and fine-tuned on a single H100 GPU.

\subsection{Baselines}
We evaluate \framework~along two dimensions: (i) \textit{How does \framework~ compare to the pretrained SwinUNETR?}, and (ii) \textit{What is the impact of pretraining on downstream performance?}

To evaluate (i), we pretrain SwinUNETR with the exact configuration given in \cite{tang2022swinunetrpret} on \brains, which includes both a contrastive loss, a rotation loss and a reconstruction loss, as well as a different choice of masking ratio and mask size. To ensure a fair comparison, SwinUNETR is pretrained for $100$ epochs, similar to \framework, and is pretrained with patch size $128^3$, instead of patch size $96^3$ and $90$ epochs. To address (ii), we apply \framework~ to a set of convolutional backbones and evaluate the difference between training from scratch and finetuning the pretrained model. The backbone models include a U-Net in two sizes: XL (90M parameters) and B. The U-Net B is very similar in size to the one used in nnUNet (22M parameters, when trained with the default setting \texttt{max\_vram} of 12 GB). Further, we explore using the modernized U-Net architecture MedNeXt \cite{roy2023mednext}, which uses depth-wise seperable convolution to introduce compound scaleable medical segmentation models. All MedNeXt models are trained with kernel size 3 and do not use UpKern. The networks are finetuned and trained from scratch using the same hyperparameters. Models trained from scratch use the full augmentation pipeline.

\begin{table}[t]
\caption{\textbf{Finetuning Results.} We evaluate \framework~ on three datasets. The numbers in {\color{red}red}/{\color{ForestGreen}green} denote benefit from pretraining. The results are Dice scores, averaged over $6$ folds. All models are trained on $n_\text{train} = 20$ samples. Standard Deviations provided in Appendix \ref{app:stddev}.}
\label{tab:results}
\resizebox{\textwidth}{!}{
\begin{tabular}{lcccllll}
\hlinebold
 & \multicolumn{2}{c}{\textbf{Pretraining}} & \textbf{Params}            & \multicolumn{1}{l}{\textbf{BraTS21}} & \multicolumn{1}{l}{\textbf{ISLES22}} & \multicolumn{1}{l}{\textbf{WMH}} & \multicolumn{1}{l}{\textbf{WMH} OOD} \\
                   & Data     &     & \textit{Millions}     & $n_\text{test} = 313$                         & $n_\text{cv} = 167$                           & $n_\text{cv} = 40$                       & $n_\text{test} = 110$                           \\ \hline
U-Net XL            & -                & \multirow{6}{*}{\rotatebox[origin=c]{90}{\small{From Scratch}}}             & 90    & $56.00$                         & $48.30$                      & $68.32$                          & $67.22$                      \\
U-Net B             & -                 &             & 22    & $55.28$                         & $45.66$                      & $66.18$                          & $64.96$                      \\
MedNeXt M \tiny{U-Net dec}        & -                 &             & 21    & $58.38$                         & $\underline{56.15}$          & $70.01$                          & $67.64$                      \\
MedNeXt L \tiny{U-Net dec}        & -                 &             & 55    & $58.28$                         & $55.40$                      & $70.32$                          & $67.57$                      \\
MedNeXt L \tiny{MedNeXt dec}        & -                 &             & 62    & $55.42$                         & $53.88$                      & $68.75$                          & $67.14$                      \\
SwinUNETR          & -                 &             & 62    & $50.00$                         & $42.55$                      & $64.96$                          & $65.02$                      \\ \hline
U-Net XL            & \brainsshort      & \framework   & 90    & $\mathbf{58.98}\dplus{2.98}$    & $53.58 \dplus{5.27}$         & $70.40 \dplus{2.08}$             & $68.22 \dplus{1.01}$         \\
U-Net B             & \brainsshort      & \framework   & 22    & $57.20\dplus{1.92}$             & $48.42 \dplus{2.76}$         & $69.11 \dplus{2.93}$             & $67.51 \dplus{2.55}$         \\
MedNeXt M         & \brainsshort      & \framework   & 21    & $56.81\dminus{1.57}$            & $55.89 \dminus{0.26}$        & $\mathbf{71.70} \dplus{1.64}$    & $\textbf{68.80} \dplus{1.17}$         \\
MedNeXt L         & \brainsshort      & \framework   & 55    & $\underline{58.73}\dplus{0.45}$ & $\textbf{56.83} \dplus{1.43}$& $\underline{71.63} \dplus{1.31}$ & $\underline{68.68} \dplus{1.11}$         \\ \hline
SwinUNETR          & \brainsshort      & \cite{tang2022swinunetrpret}   & 62    & $49.54\dminus{0.46}$            & $48.42 \dplus{1.75}$         & $67.28 \dplus{2.32}$             & $68.14 \dplus{1.59}$         \\ \hlinebold
\end{tabular}
}
\end{table}

\section{Results and Discussion}
\label{sec:results}
The results of our experiments are given in Table \ref{tab:results}. Both U-Net B, U-Net XL and MedNeXt models benefit from pretraining in all scenarios. MedNeXt M does not benefit from pretraining on BraTS21 and ISLES22, however exhibits very good performance on WMH both in and out of domain. We believe this is mainly due to strong baseline performance, before pretraining is added. Note that models do not use the full augementation pipeline during finetuning, while the baselines from scratch use all augmentations. All models trained with \framework~ outperform the pretrained SwinUNETR, except for the standard U-Net on WMH OOD. Generally, model size seems to be correlated with performance, with two notable exceptions: The SwinUNETR model seems to be worse than both MedNeXt M model and the standard U-Net, both of which have less than a third of the parameters of SwinUNETR. The results on WMH OOD demonstrate that pretraining also increases performance when applied to out-of-domain data -- even with respect to the pretraining domain.

\textbf{Limitations}. This study only considers single sequence and low resource settings. We have not applied \framework~ to the multi-sequence versions of the dataset and neither have we applied it to datasets with more datapoints than $n_{\text{train}} = 20$. How to leverage pretraining for multi-sequence tasks is a relevant avenue for future work. This study neither tests the SwinUNETR framework using different backbones and nor tests \framework~ using a SwinUNETR backbone, since it is not readily compatible with a U-Net decoder.\footnote{Since it uses non-pretrained UNETR \cite{Hatamizadeh2021UNETR} encoder blocks in the interface between the Swin-Transformer and the convolutional decoder.}

\section{Conclusion}
We present \framework, a framework based on Masked Image Modeling to pretrain 3D-native segmentation models which balances memory usage, runtime, and finetuning performance. We further introduce \brains, the largest dataset of brain MRI for pretraining to date. We pretrain \framework~ on \brains~ and evaluate the models on three challenging datasets. We compare the framework to models trained from scratch as well as the pretrained SwinUNETR model. Our results demonstrate that when \framework~is used to pretrain U-Net models or MedNeXt models, the models outperform similar models trained from scratch, as well as SwinUNETR models, both pretrained and not in the challenging tasks evaluated.
\\

\noindent\textbf{Acknowledgements}. This work was supported by Danish Data Science Academy, which is funded by the Novo Nordisk Foundation
(NNF21SA0069429) and Villum Fonden (40516), Digital Research Centre Denmark with grant number 9142-00001B, and Pioneer Centre for AI, Danish National Research Foundation, grant number P1.

\bibliography{main}

\begin{thebibliography}{10}
\providecommand{\url}[1]{\texttt{#1}}
\providecommand{\urlprefix}{URL }
\providecommand{\doi}[1]{https://doi.org/#1}

\bibitem{bao2021beit}
Bao, H., Dong, L., Piao, S., Wei, F.: Beit: Bert pre-training of image transformers. In: International Conference on Learning Representations (2021)

\bibitem{chaitanya2020contrastive}
Chaitanya, K., Erdil, E., Karani, N., Konukoglu, E.: Contrastive learning of global and local features for medical image segmentation with limited annotations. Advances in neural information processing systems  \textbf{33},  12546--12558 (2020)

\bibitem{chen2020simple}
Chen, T., Kornblith, S., Norouzi, M., Hinton, G.: A simple framework for contrastive learning of visual representations. In: International conference on machine learning. pp. 1597--1607. PMLR (2020)

\bibitem{chlap2021review}
Chlap, P., Min, H., Vandenberg, N., Dowling, J., Holloway, L., Haworth, A.: A review of medical image data augmentation techniques for deep learning applications. Journal of Medical Imaging and Radiation Oncology  \textbf{65}(5),  545--563 (2021)

\bibitem{msseg1}
Commowick, O., Istace, A., Kain, M., Laurent, B., Leray, F., Simon, M., Camarasu-Pop, S., Girard, P., Ameli, R., Ferré, J.C., Kerbrat, A., Tourdias, T., Cervenansky, F., Glatard, T., Beaumont, J., Doyle, S., Forbes, F., Knight, J., Khademi, A., Barillot, C.: Objective evaluation of multiple sclerosis lesion segmentation using a data management and processing infrastructure. Scientific Reports  \textbf{8},  13650--13666 (09 2018). \doi{10.1038/s41598-018-31911-7}

\bibitem{grill2020bootstrap}
Grill, J.B., Strub, F., Altch{\'e}, F., Tallec, C., Richemond, P., Buchatskaya, E., Doersch, C., Avila~Pires, B., Guo, Z., Gheshlaghi~Azar, M., et~al.: Bootstrap your own latent-a new approach to self-supervised learning. Advances in neural information processing systems  \textbf{33},  21271--21284 (2020)

\bibitem{brats21}
Guzmán Pérez-Carrillo, G.: The rsna-asnr-miccai brats 2021 benchmark on brain tumor segmentation and radiogenomic classification  (09 2021)

\bibitem{Hatamizadeh2021UNETR}
Hatamizadeh, A., Yang, D., Roth, H.R., Xu, D.: Unetr: Transformers for 3d medical image segmentation. 2022 IEEE/CVF Winter Conference on Applications of Computer Vision (WACV) pp. 1748--1758 (2021), \url{https://api.semanticscholar.org/CorpusID:232290634}

\bibitem{he2022masked}
He, K., Chen, X., Xie, S., Li, Y., Doll{\'a}r, P., Girshick, R.: Masked autoencoders are scalable vision learners. In: Proceedings of the IEEE/CVF conference on computer vision and pattern recognition. pp. 16000--16009 (2022)

\bibitem{Isensee2020nnUNet}
Isensee, F., Jaeger, P.F., Kohl, S.A.A., Petersen, J., Maier-Hein, K.: nnu-net: a self-configuring method for deep learning-based biomedical image segmentation. Nature Methods  \textbf{18},  203 -- 211 (2020), \url{https://www.nature.com/articles/s41592-020-01008-z}

\bibitem{isensee2024nnu}
Isensee, F., Wald, T., Ulrich, C., Baumgartner, M., Roy, S., Maier-Hein, K., Jaeger, P.F.: nnu-net revisited: A call for rigorous validation in 3d medical image segmentation. arXiv preprint arXiv:2404.09556  (2024)

\bibitem{Isensee2024nnUnetRevisited}
Isensee, F., Wald, T., Ulrich, C., Baumgartner, M., Roy, S., Maier-Hein, K., Jaeger, P.F.: nnu-net revisited: A call for rigorous validation in 3d medical image segmentation. ArXiv Preprint arXiv:2404.09556  (2024)

\bibitem{oasis4}
Koenig, L.N., Day, G.S., Salter, A., Keefe, S., Marple, L.M., Long, J., LaMontagne, P., Massoumzadeh, P., Snider, B.J., Kanthamneni, M., Raji, C.A., Ghoshal, N., Gordon, B.A., Miller-Thomas, M., Morris, J.C., Shimony, J.S., Benzinger, T.L.: Select atrophied regions in alzheimer disease (sara): An improved volumetric model for identifying alzheimer disease dementia. NeuroImage: Clinical  \textbf{26},  102248 (2020). \doi{https://doi.org/10.1016/j.nicl.2020.102248}, \url{https://www.sciencedirect.com/science/article/pii/S2213158220300851}

\bibitem{wmh}
Kuijf, H., Biesbroek, M., de~Bresser, J., Heinen, R., Andermatt, S., Bento, M., Berseth, M., Belyaev, M., Cardoso, M.J., Casamitjana, A., Collins, L., Dadar, M., Georgiou, A., Ghafoorian, M., Jin, D., Khademi, A., Knight, J., Li, H., Llado, X., Biessels, G.: Standardized assessment of automatic segmentation of white matter hyperintensities; results of the wmh segmentation challenge. IEEE Transactions on Medical Imaging  \textbf{PP}, ~1--1 (03 2019). \doi{10.1109/TMI.2019.2905770}

\bibitem{oasis3}
LaMontagne, P., Benzinger, T., Morris, J., Keefe, S., Hornbeck, R., Xiong, C., Grant, E., Hassenstab, J., Moulder, K., Vlassenko, A., Raichle, M., Cruchaga, C., Marcus, D.: Oasis-3: Longitudinal neuroimaging, clinical, and cognitive dataset for normal aging and alzheimer disease  (12 2019). \doi{10.1101/2019.12.13.19014902}

\bibitem{llambias2024yucca}
Llambias, S.N., Machnio, J., Munk, A., Ambsdorf, J., Nielsen, M., Ghazi, M.M.: Yucca: A deep learning framework for medical image analysis. arXiv preprint arXiv:2407.19888  (2024), \url{https://arxiv.org/abs/2407.19888}

\bibitem{ppmi}
Marek, K., et. al.: The parkinson progression marker initiative (ppmi). Progress in Neurobiology  \textbf{95}(4),  629--635 (2011). \doi{https://doi.org/10.1016/j.pneurobio.2011.09.005}, \url{https://www.sciencedirect.com/science/article/pii/S0301008211001651}

\bibitem{noroozi2016unsupervised}
Noroozi, M., Favaro, P.: Unsupervised learning of visual representations by solving jigsaw puzzles. In: European conference on computer vision. pp. 69--84. Springer (2016)

\bibitem{adni}
Peters, R.C., et. al.: Alzheimer's disease neuroimaging initiative (adni): Clinical characterization. Neurology  \textbf{74}(3),  201--209 (Jan 2010). \doi{10.1212/WNL.0b013e3181cb3e25}

\bibitem{isles22}
Petzsche, M., de~la Rosa, E., Hanning, U., Wiest, R., Valenzuela, W., Reyes, M., Meyer, M., Liew, S.L., Kofler, F., Ezhov, I., Robben, D., Hutton, A., Friedrich, T., Zarth, T., Bürkle, J., Baran, T., Menze, B., Broocks, G., Meyer, L., Kirschke, J.: Isles 2022: A multi-center magnetic resonance imaging stroke lesion segmentation dataset. Scientific Data  \textbf{9}, ~762 (12 2022). \doi{10.1038/s41597-022-01875-5}

\bibitem{roy2023mednext}
Roy, S., Koehler, G., Ulrich, C., Baumgartner, M., Petersen, J., Isensee, F., Jaeger, P.F., Maier-Hein, K.H.: Mednext: transformer-driven scaling of convnets for medical image segmentation. In: International Conference on Medical Image Computing and Computer-Assisted Intervention. pp. 405--415. Springer (2023)

\bibitem{msd}
Simpson, A., Antonelli, M., Bakas, S., Bilello, M., Farahani, K., Ginneken, B., Kopp-Schneider, A., Landman, B., Litjens, G., Menze, B., Ronneberger, O., Summers, R., Bilic, P., Christ, P., Do, R., Gollub, M., Golia-Pernicka, J., Heckers, S., Jarnagin, W., Cardoso, M.J.: A large annotated medical image dataset for the development and evaluation of segmentation algorithms  (02 2019)

\bibitem{tang2022swinunetrpret}
Tang, Y., Yang, D., Li, W., Roth, H.R., Landman, B., Xu, D., Nath, V., Hatamizadeh, A.: Self-supervised pre-training of swin transformers for 3d medical image analysis. In: Proceedings of the IEEE/CVF Conference on Computer Vision and Pattern Recognition. pp. 20730--20740 (2022)

\bibitem{vincent2008extracting}
Vincent, P., Larochelle, H., Bengio, Y., Manzagol, P.A.: Extracting and composing robust features with denoising autoencoders. In: Proceedings of the 25th international conference on Machine learning. pp. 1096--1103 (2008)

\bibitem{Woo2023ConvNeXtV2}
Woo, S., Debnath, S., Hu, R., Chen, X., Liu, Z., Kweon, I.S., Xie, S.: Convnext v2: Co-designing and scaling convnets with masked autoencoders pp. 16133--16142 (June 2023)

\bibitem{zhang2016colorful}
Zhang, R., Isola, P., Efros, A.A.: Colorful image colorization. In: Computer Vision--ECCV 2016: 14th European Conference, Amsterdam, The Netherlands, October 11-14, 2016, Proceedings, Part III 14. pp. 649--666. Springer (2016)

\end{thebibliography}

\newpage

\appendix
\setcounter{page}{1}
{\centering \Large \textbf{Supplementary Material}}

\section{Hyperparameters}
\label{app:hyperparameters}
\begin{table}[H]
\scriptsize
\begin{tabular}{l|cccccccccc}
                        & \multicolumn{5}{c}{Pretrain}                                        & \multicolumn{5}{c}{Finetune/From scratch}                           \\
                        & \multicolumn{2}{c}{U-Net} & \multicolumn{2}{c}{MedNeXt} & SwinUNETR & \multicolumn{2}{c}{U-Net} & \multicolumn{2}{l}{MedNeXt} & SwinUNETR \\
Config                  & XL           &  B         & L           & M           &  -         & XL          &  B           & L           & M           &      -     \\ \hline
optimizer               & \multicolumn{10}{c}{AdamW}                                                                                                                \\
base learning rate      & \multicolumn{5}{c}{$1e-4$}                                           & $1e-4$      & $1e-4$      & $5e-4$       & $6e-4$       & $1e-4$    \\
weight decay            & \multicolumn{10}{c}{$3e-5$}                                                                                                               \\
batch size              & 16           & 16         & 8            & 8            & 2         & \multicolumn{5}{c}{4}                                               \\
learning rate scheduler & \multicolumn{10}{c}{cosine decay}                                                                                                         \\
warmup epochs           & 5            & 5          & 5            & 5            & 5         & 0           & 0           & 0            & 0            & 0         \\
amsgrad                 & \multicolumn{10}{c}{True}                      \\
optimizer eps           & \multicolumn{10}{c}{$1e-8$}                                                                                                               \\
precision               & \multicolumn{10}{c}{bf16-mixed}                                                                                                           \\
compile                 & yes          & yes        & yes          & yes          & no        & yes         & yes         & yes          & yes          & no        \\
mask ratio              & 0.6          & 0.6        & 0.6          & 0.6          & 0.3       & -           & -           & -            & -            & -         \\
mask size               & \multicolumn{4}{c}{$4^3$}                               & 2         & -           & -           & -            & -            & -         \\
patch size              & \multicolumn{10}{c}{$128^3$}                                                                                 
\end{tabular}
\end{table}

\section{Augmentations}

\label{app:augmentations}
\begin{table}[H]
\centering
\caption{\textbf{Augmentations}. {$p_{\text{sample}}$} denotes the probability a given augmentation is applied to the sample. {$p_{\text{axis}}$} denotes the probability the given augmentation is applied to each axis, if relevant.}
\label{tab:augs}
\begin{tabular}{clccl}
\hlinebold
& Augmentation                    & {$p_{\text{sample}}$} & {$p_{\text{axis}}$} & Notes \\ \hline
\multirow{2}{*}{\rotatebox[origin=c]{90}{\tiny{\textit{Spatial}}}}
& Rotation                        & 0.2                    & 0.66        & Rotation in $[-30,30]$ for all axis  \\
& Scale                           & 0.2                    & -           &  Scaling factor in $[0.9,1.1]$        \\ \hline
\multirow{9}{*}{\rotatebox[origin=c]{90}{\tiny{\textit{Intensity}}}} 
& Elastic Deformation            & 0.33                    & -           & $\alpha \in [200,600], \sigma \in [20,30]$           \\
& Gaussian Blur                   & 0.1                    & -           &  $\mu = 0, \sigma \in [0,1]$        \\
& Gaussian Additive Noise         & 0.2                    & -           & $\mu = 0, \sigma \in [1^{-3}, 1^{-4}]$\\
& Gaussian Multiplicative Noise   & 0.2                    & -           & $\mu = 0, \sigma \in [1^{-3}, 1^{-4}]$            \\
& Gamma                           & 0.2                    & -           &  Inverts colors of image with $p = 0.01$   \\
& Motion Ghosting                 & 0.2                    & -           &  $\alpha \in [0.85, 0.95]$, $\text{repetitions} \in [2, 11]$    \\
& Biasfield                       & 0.33                   & -           &   -     \\
& Gibbs Ringing                   & 0.2                    & -           &   -     \\
& Simulate Low Resolution \:\:    & 0.1                    & 0.33        &   Zoom range in $[0.5, 1.0]$    \\ \hlinebold
\end{tabular}
\end{table}

\section{ADNI Acknowledgement}
\tiny{
Data used in preparation of this article were obtained from the Alzheimer’s Disease Neuroimag- ing Initiative (ADNI) database (adni.loni.usc.edu). As such, the investigators within the ADNI contributed to the design and implementation of ADNI and/or provided data but did not par- ticipate in analysis or writing of this report. A complete listing of ADNI investigators can be found 
\href{http://adni.loni.usc.edu/wp-content/uploads/how_to_apply/ ADNI_Acknowledgement_List.pdf}{here}.

The ADNI was launched in 2003 as a public-private partnership, led by Principal Investigator Michael W. Weiner,MD. The primary goal of ADNI has been to test whether serial magnetic reso- nance imaging (MRI), positron emission tomography (PET), other biological markers, and clinical and neuropsychological assessment can be combined to measure the progression of mild cognitive impairment (MCI) and early Alzheimer’s disease (AD). For up-to-date information, see \url{www.adni-info.org}.

Data collection and sharing for this project was funded by the Alzheimer’s Disease Neuroimaging Initiative (ADNI) (National Institutes of Health Grant U01 AG024904) and DOD ADNI (Depart- ment of Defense award number W81XWH-12-2-0012). ADNI is funded by the National Institute on Aging, the National Institute of Biomedical Imaging and Bioengineering, and through generous contributions from the following: AbbVie, Alzheimer’s Association; Alzheimer’s Drug Discovery Foundation; Araclon Biotech; BioClinica, Inc.; Biogen; Bristol-Myers Squibb Company; CereSpir, Inc.; Cogstate; Eisai Inc.; Elan Pharmaceuticals, Inc.; Eli Lilly and Company; EuroImmun; F. Hoffmann-La Roche Ltd and its affiliated company Genentech, Inc.; Fujirebio; GE Healthcare; IXICO Ltd.;Janssen Alzheimer Immunotherapy Research \& Development, LLC.; Johnson \& John- son Pharmaceutical Research \& Development LLC.; Lumosity; Lundbeck; Merck \& Co., Inc.;Meso Scale Diagnostics, LLC.; NeuroRx Research; Neurotrack Technologies; Novartis Pharmaceuticals Corporation; Pfizer Inc.; Piramal Imaging; Servier; Takeda Pharmaceutical Company; and Transi- tion Therapeutics. The Canadian Institutes of Health Research is providing funds to support ADNI clinical sites in Canada. Private sector contributions are facilitated by the Foundation for the National Institutes of Health (www.fnih.org). The grantee organization is the Northern California Institute for Research and Education, and the study is coordinated by the Alzheimer’s Therapeutic Research In- stitute at the University of Southern California. ADNI data are disseminated by the Laboratory for Neuro Imaging at the University of Southern California.}

\section{Contrastive and Rotation Loss Curves}
\label{app:contrastive-rotation}
\begin{figure}
     \centering
     \begin{subfigure}[b]{0.49\textwidth}
         \centering
         \includegraphics[width=\textwidth]{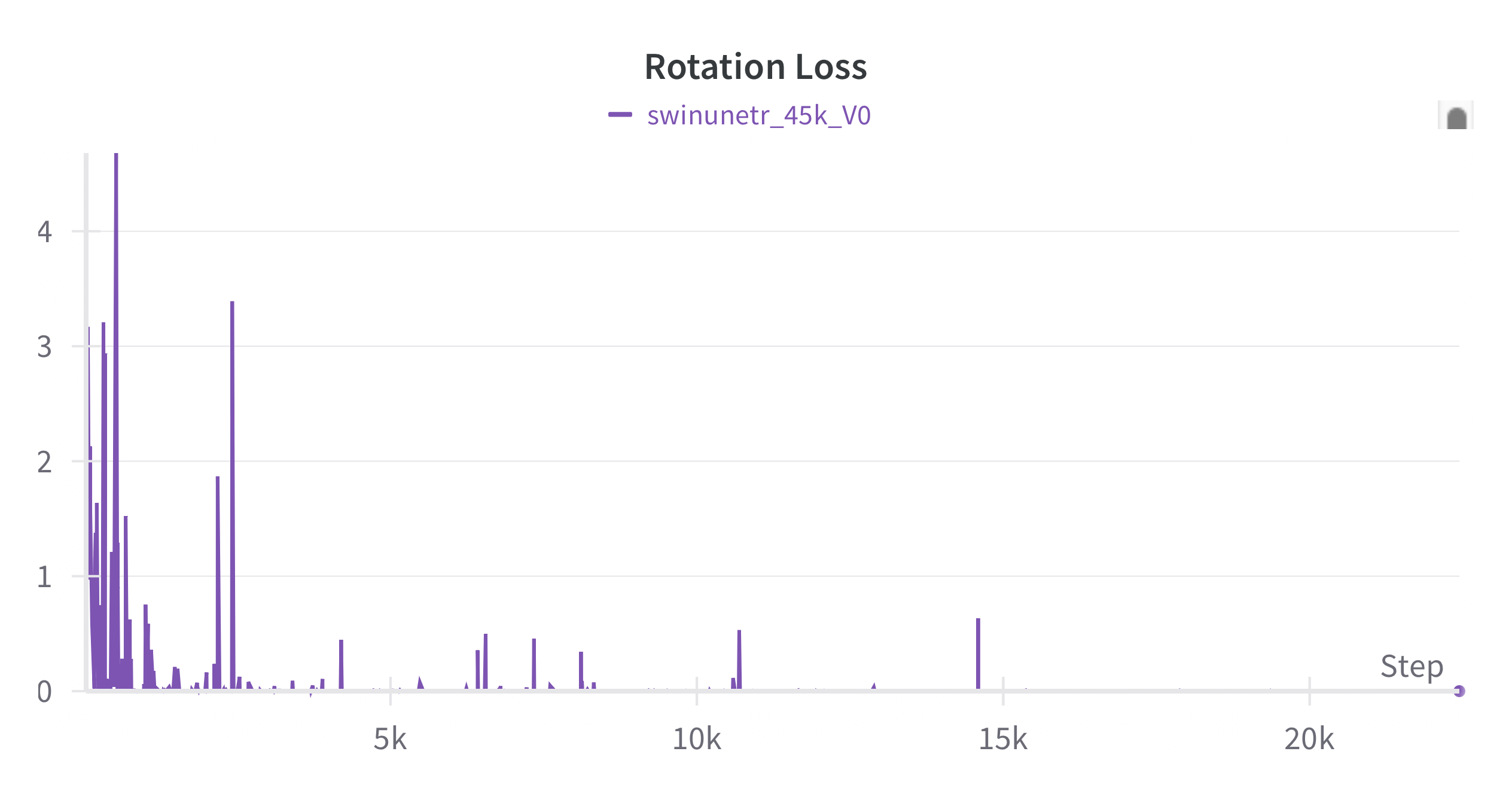}
         \caption{Rotation loss}
         \label{fig:rotation}
     \end{subfigure}
     \hfill
     \begin{subfigure}[b]{0.49\textwidth}
         \centering
         \includegraphics[width=\textwidth]{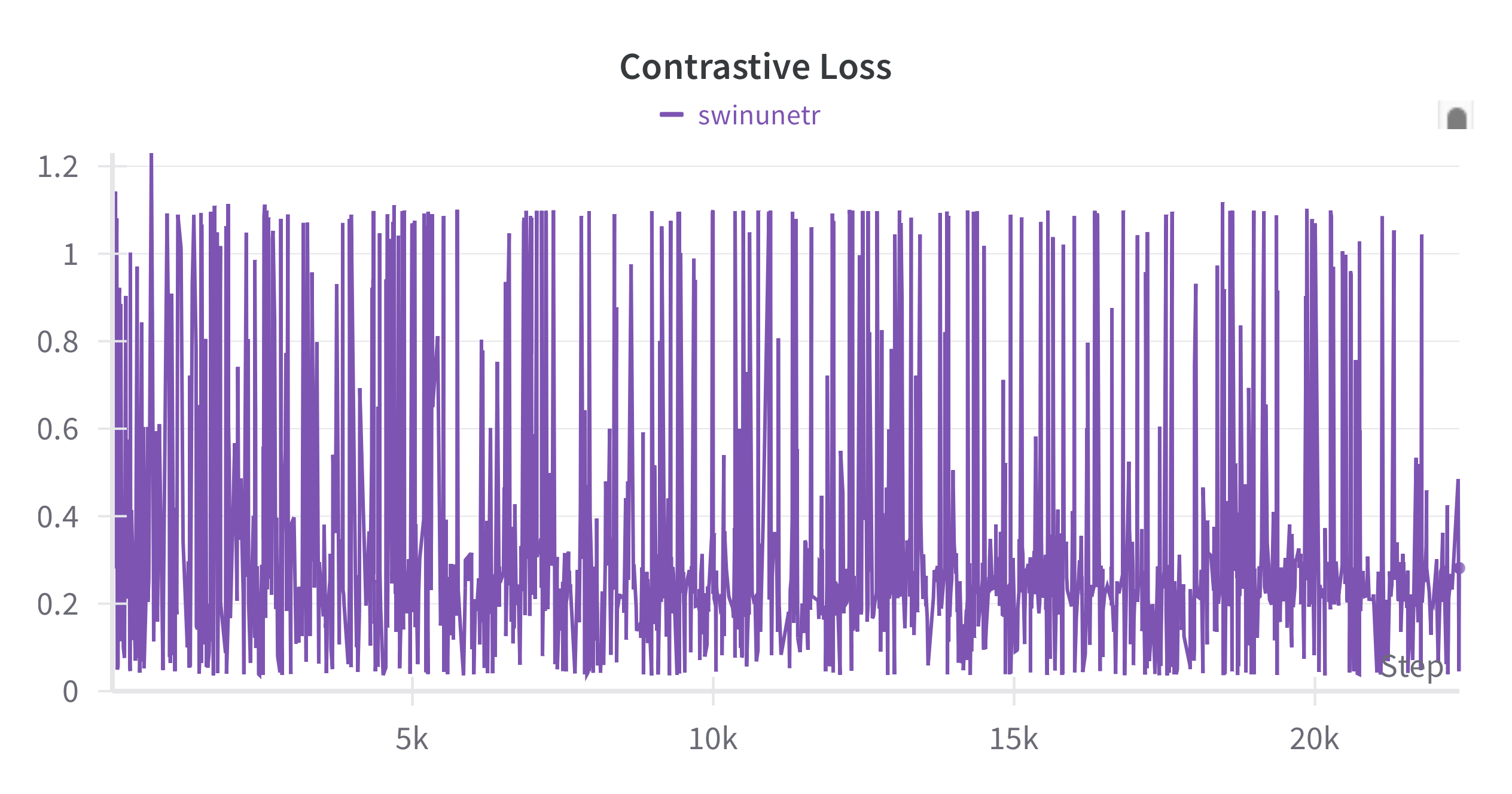}
         \caption{Contrastive Loss}
         \label{fig:contrastive}
     \end{subfigure}
    \caption{Rotation and contrastive losses when training SwinUNETR \cite{tang2022swinunetrpret} on \brains.}
    \label{fig:rot_cont_losses}
\end{figure}

\section{Results with standard deviation}
\label{app:stddev}
\begin{table}[H]
\caption{Main results (as Table \ref{tab:results}) with variation at one standard deviation.}
\resizebox{\textwidth}{!}{
\begin{tabular}{lcccllll}
\hlinebold
 & \multicolumn{2}{c}{\textbf{Pretraining}} & \textbf{Params}    & \multicolumn{1}{l}{\textbf{BraTS21}} & \multicolumn{1}{l}{\textbf{ISLES22}} & \multicolumn{1}{l}{\textbf{WMH}} & \multicolumn{1}{l}{\textbf{WMH} OOD} \\
                   & Data             &       & \textit{Millions}     & $n_\text{test} = 313$                         & $n_\text{cv} = 167$                           & $n_\text{cv} = 40$                       & $n_\text{test} = 110$                           \\ \hline
U-Net XL                              & -     & \multirow{6}{*}{\rotatebox[origin=c]{90}{\small{From Scratch}}}            
                                                     & 90    & $56.00 \pm 2.55$     & $48.30 \pm 2.94$                      & $68.32 \pm 1.75$                          & $67.22 \pm 1.59$                      \\
U-Net B                               & -     &      & 22    & $55.28 \pm 1.75$     & $45.66 \pm 2.40$                      & $66.18 \pm 2.28$                          & $64.96 \pm 2.52$                      \\
MedNeXt M \tiny{U-Net dec}            & -     &      & 21    & $58.38 \pm 1.44$     & $56.15 \pm 2.77$                      & $70.01 \pm 1.48$                          & $67.64 \pm 1.74$                      \\
MedNeXt L \tiny{U-Net dec}            & -     &      & 55    & $58.28 \pm 1.23$     & $55.40 \pm 1.88$                      & $70.32 \pm 1.77$                          & $67.57 \pm 1.48$                      \\
MedNeXt L \tiny{MedNeXt dec}          & -     &      & 62    & $55.42 \pm 2.25$     & $53.88 \pm 1.52$                      & $68.75 \pm 2.18$                          & $67.14 \pm 1.28$                      \\
SwinUNETR          & -                &              & 62    & $50.00 \pm 2.59$     & $42.55 \pm 2.49$                      & $64.96 \pm 3.90$                          & $65.02 \pm 2.86$                      \\ \hline
U-Net XL            & \brainsshort    & \framework   & 90    & $58.98 \pm 0.88$     & $53.58 \pm 0.76$                      & $70.40 \pm 1.46$                          & $68.22 \pm 1.16$                       \\
U-Net B             & \brainsshort    & \framework   & 22    & $57.20 \pm 1.75$     & $48.42 \pm 1.32$                      & $69.11 \pm 2.28$                          & $67.51 \pm 1.56$                       \\
MedNeXt M         & \brainsshort      & \framework   & 21    & $56.81 \pm 2.40$     & $55.89 \pm 2.17$                      & $71.70 \pm 1.62$                          & $68.80 \pm 0.79$                       \\
MedNeXt L         & \brainsshort      & \framework   & 55    & $58.73 \pm 1.33$     & $56.83 \pm 1.39$                      & $71.63 \pm 1.67$                          & $68.68 \pm 0.97$                       \\ \hline
SwinUNETR          & \brainsshort     & \cite{tang2022swinunetrpret}   & 62    & $49.54 \pm 0.94$ & $48.42 \pm 3.14$        & $67.28 \pm 2.67$                          & $68.14 \pm 1.43$                       \\ \hlinebold
\end{tabular}
}
\end{table}

\end{document}